\documentclass{aa}

\usepackage{graphicx}
\usepackage{natbib}

\bibpunct{(}{)}{;}{a}{}{,}
\usepackage{txfonts}

\begin{document}
\title{Chemistry and line emission from evolving Herbig Ae disks}

\author{B. Jonkheid\inst{1} \and C.P. Dullemond\inst{2} \and M.R. Hogerheijde\inst{1} \and E.F. van Dishoeck\inst{1}}

\authorrunning{Jonkheid et al.}
\titlerunning{}

\offprints{B. Jonkheid, \email{jonkheid@strw.leidenuniv.nl}}

\institute{
Sterrewacht Leiden, Leiden University, P.O. Box 9513, 2300 RA Leiden, the Netherlands
\and
Max Planck Institut f\"ur Astronomie, K\"onigstuhl 17, 69117 Heidelberg, Germany
}

\date{Received / Accepted }

\abstract{}{
To calculate chemistry and gas temperature of evolving protoplanetary disks 
with decreasing mass or dust settling, and to explore the sensitivity of 
gas-phase tracers.
}{
The density and dust temperature profiles for a range of models of flaring and 
self-shadowed disks around a typical Herbig Ae star are used together with 
2-dimensional ultraviolet (UV) radiative transfer to calculate the chemistry 
and gas temperature. In each model the line profiles and intensities for the 
fine structure lines of [\ion{O}{i}], [\ion{C}{ii}] and 
[\ion{C}{i}] and the pure rotational lines of CO, CN, HCN and ${\rm HCO^+}$ are determined.
}{
The chemistry shows a strong correlation with disk mass. Molecules that are
easily dissociated, like HCN, require high densities and large extinctions
before they can become abundant. The products of photodissociation, like 
CN and ${\rm C_2H}$, become abundant in models with lower masses.
Dust settling mainly affects the gas temperature, and thus high temperature 
tracers like the O and ${\rm C^+}$ fine structure lines. The carbon chemistry 
is found to be very sensitive to the adopted PAH abundance. The line ratios 
CO/${\rm ^{13}CO}$, CO/${\rm HCO^+}$ and [\ion{O}{i}] ${\rm 63\,\mu m}$/${\rm 146\,\mu m}$ can be used to 
distinguish between disks where dust growth and settling takes place, and disks
that undergo overall mass loss.
}{}


\maketitle

\section{Introduction}
Circumstellar disks are a natural by-product of the formation of stars of low 
to intermediate masses, the former objects being identified as T-Tauri stars 
while the latter are Herbig Ae/Be stars. From models of the spectral energy 
distributions (SEDs) at infrared and millimeter wavelengths, which track the 
dust emission from these disks, it has been concluded that some disks have a 
flaring geometry, allowing them to intercept a significant portion of the 
stellar radiation and re-emitting it at longer wavelengths 
\citep[e.g.][]{kenhar87,chigol97}. In a recent study, \citet{duldom04} 
suggested that the difference between flaring (`group I') and flat (`group
II') disks can be explained by the presence of a puffed-up inner rim that casts
a shadow over the disk. Depending on whether there is enough material in the 
outer disk to rise above the shadow either an illuminated flared disk 
or a self-shadowed flat disk results. One interpretation of the occurrence of
group I and group II sources is that flared disks evolve into flat disks when
the dust grows and settles to the midplane, or when the disk's mass decreases
due to dissipation and evaporation.

\begin{figure*}[!tp]
\centering
\includegraphics[width=17cm]{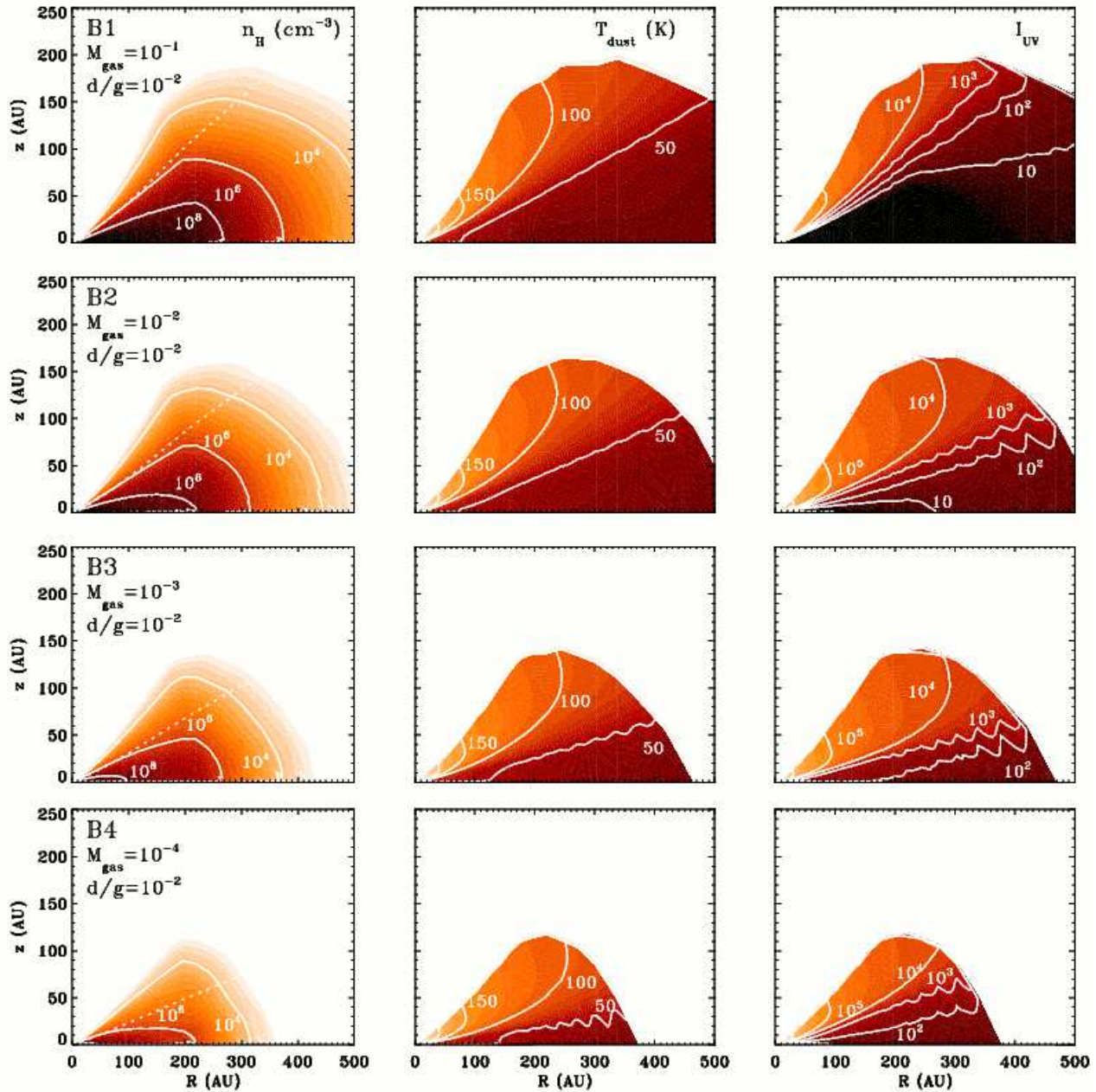}
\caption{Input densities (left column) dust temperatures (middle column) and integrated strength of the UV field (right column) for models B1--B4 of \citet{duldom04}. The gas mass (in $M_\odot$) and dust/gas ratio are indicated in the for each row. The UV field is given in terms of $I_{\rm UV}$ times the strength of the interstellar radiation field by \citet{draine78} integrated between 912 and 2050 \AA. The dotted lines in the density plots give the boundary where the visual extinction towards the central star is 1 magnitude.}
\label{dd04b}
\end{figure*}

\begin{figure*}[!tp]
\centering
\includegraphics[width=17cm]{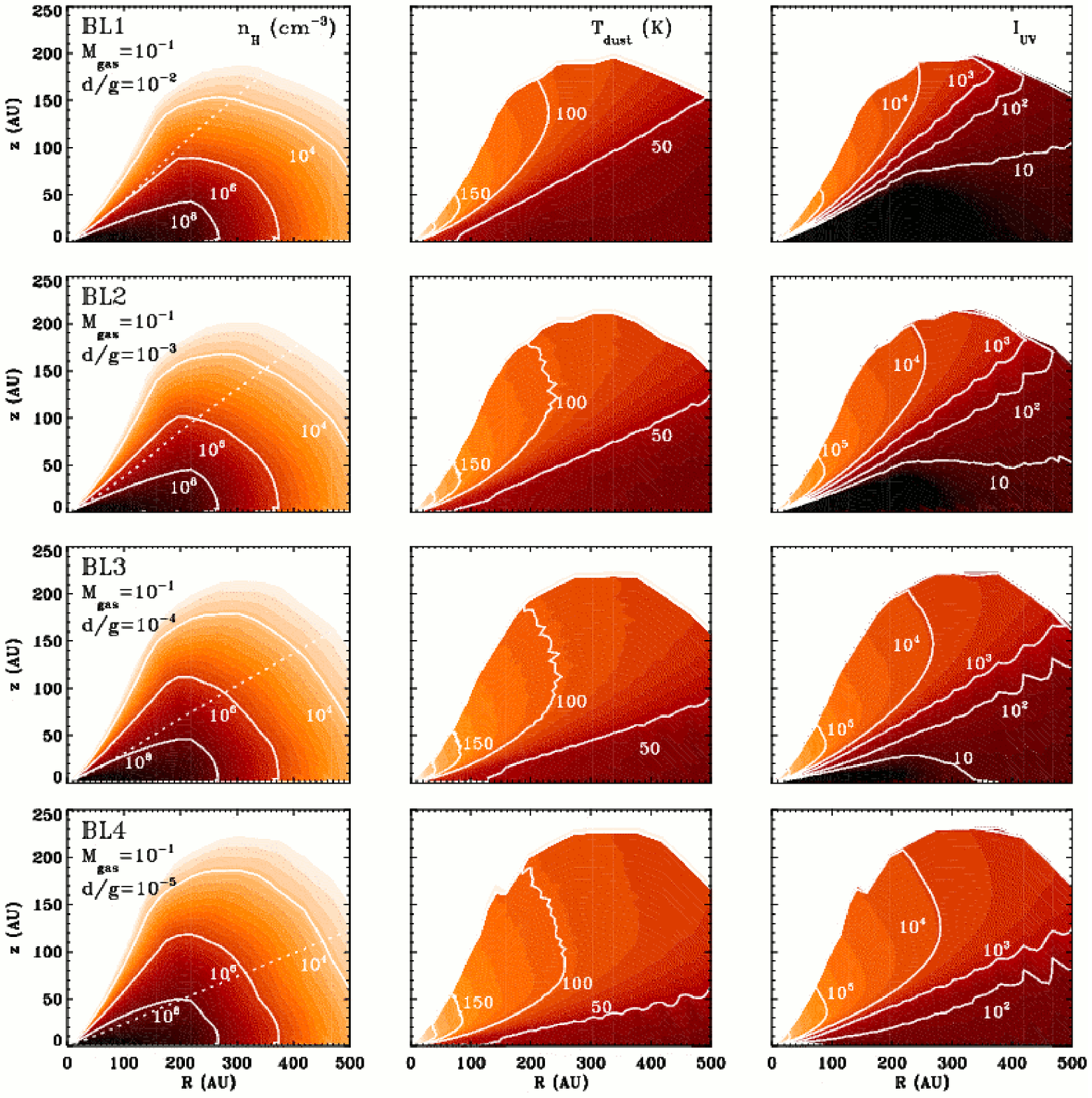}
\caption{As Figure \ref{dd04b}, but for models BL1--BL4.}
\label{dd04bl}
\end{figure*}

While SEDs are useful to examine the properties of the dust content of disks, 
the physical and chemical properties of the gas are studied through line 
emission from both fine-structure transitions in atoms and rotational and 
vibrational transitions in molecules. The pure rotational lines and 
ro-vibrational lines of CO 
\citep[e.g.][]{koesar95,dutrey97,thi01,britta03,dent05}, and ${\rm H_2}$ 
\citep[e.g.][]{bary03} have been observed in disks, as well as the 
rotational lines of several other molecules like CN, HCN, ${\rm HCO^+}$ and 
${\rm C_2H}$ \citep[e.g.][]{dutrey97,thi04}. To obtain information on the
temperature, chemical composition and total mass of the gas
from the observed emission lines, the chemistry of the disk has to be modelled.
In recent years calculations of both the chemistry 
\citep[e.g.][]{aikawa97,zadelh03} and the gas temperature 
\citep[e.g.][]{jonkhe04,kamdul04} have been performed using a density
structure based on previous models of the disk structure 
\citep[e.g.][]{daless98,dullem02}. In these models, the density structure is 
calculated assuming vertical hydrostatic equilibrium, and the dust temperature
is obtained by solving the radiative transfer of stellar light. In most cases 
the gas temperature is assumed to be equal to the dust temperature, but some
models calculate the gas temperature and density structure in equilibrium with 
each other \citep[e.g.][]{gorhol04,nommil05}. The emission lines are modelled 
from the results of the chemistry and temperature calculations 
\citep[e.g.][]{kamp03,jonkhe04,gorhol04}. Effects of grain growth and dust 
settling on the chemistry have also been considered 
\citep[e.g.][]{kamp03,jonkhe04,jonkhe06,aiknom06}.

In the current paper calculations of the chemistry and gas temperature are 
presented based on the disk structure models of group I and group II sources
by \citet{duldom04} to investigate how the chemistry and line emisison 
change in a series of evolving disks. Special
attention was given to identify possible tracers of the disk geometry that can 
supplement and confirm the information gained from SEDs. This work 
complements the calculations for optically thin disks by \citet{kamp06}.

\section{Model}

\begin{figure}
\resizebox{\hsize}{!}{\includegraphics[angle=0]{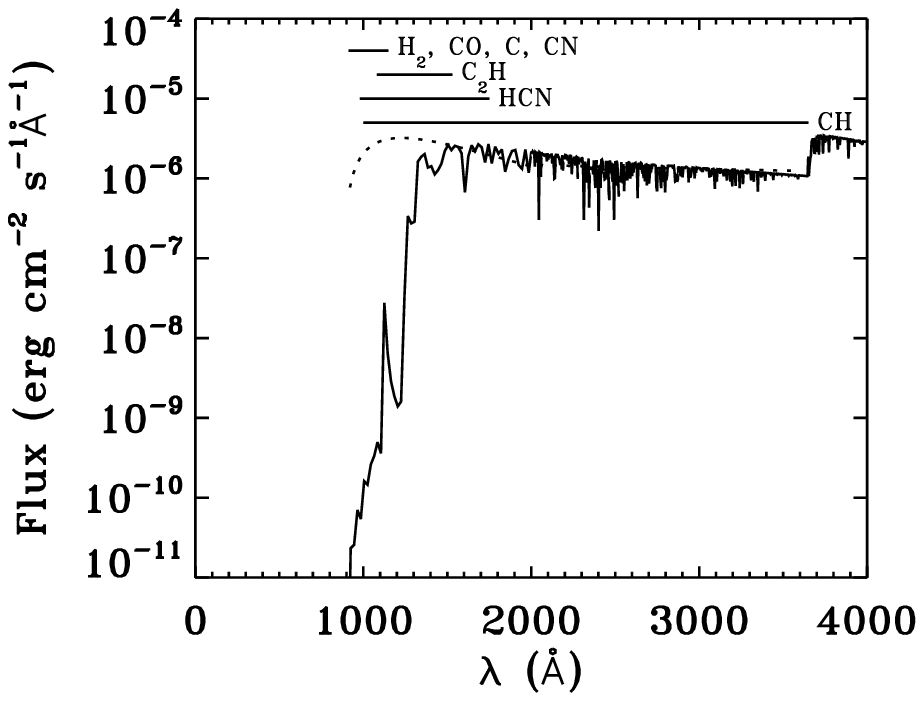}}
\caption{The radiation fields used in the model. The solid line gives the stellar radiation field, scaled to the interstellar UV field by \citet{draine78} (dotted line). The horizontal lines indicate the wavelength ranges where several important molecules are dissociated.}
\label{starspec}
\end{figure}

\begin{table}
\centering
\caption[]{Model parameters used in the calculations. Note that the dust masses given here include only small (interstellar sized) grains.}
\label{models}
\begin{tabular}{l l l}
\hline\hline
Model & $M_{\rm gas}\,(M_\odot)$ & $M_{\rm dust}\,(M_\odot)$\\
\hline
B1 & $10^{-1}$ & $10^{-3}$\\
B2 & $10^{-2}$ & $10^{-4}$\\
B3 & $10^{-3}$ & $10^{-5}$\\
B4 & $10^{-4}$ & $10^{-6}$\\
\hline
BL1 & $10^{-1}$ & $10^{-3}$\\
BL2 & $10^{-1}$ & $10^{-4}$\\
BL3 & $10^{-1}$ & $10^{-5}$\\
BL4 & $10^{-1}$ & $10^{-6}$\\

\hline
\end{tabular}
\end{table}

\begin{table}
\centering
\caption[]{Adopted gas-phase elemental abundances with respect to total hydrogen.}
\label{elabun}
\begin{tabular}{l l}
\hline\hline
Element & abundance\\
\hline
C & $1.2\times 10^{-4}$\\
O & $1.6\times 10^{-4}$\\
N & $2.0\times 10^{-5}$\\
Mg & $4.2\times 10^{-6}$\\
S & $1.9\times 10^{-6}$\\
Si & $8.0\times 10^{-7}$\\
Fe & $4.3\times 10^{-6}$\\
PAH & $ 1\times10^{-10}\,-\,1\times 10^{-7\ {\rm a}}$\\
\hline
\end{tabular}
\begin{enumerate}
\item[${\rm ^a}$] In the B series of models, the PAH abundance is kept constant at $10^{-7}$. In the BL series, the PAH abundance is lowered together with the dust content of the disk, going from $10^{-7}$ in BL1 to $10^{-10}$ in BL4.
\end{enumerate}
\end{table}

The chemistry is calculated for a range of disk structure models by 
\citet{duldom04} in which the disk mass and dust/gas ratio are varied 
systematically (see Table \ref{models} and Figures \ref{dd04b} and 
\ref{dd04bl}). In that paper the vertical structure of the disk is computed in 
hydrostatic equilibrium assuming several radial distributions and total disk 
masses. In all models the gas temperature was taken equal to the dust 
temperature. Under consideration here are models B1--B4
and BL1--BL4. These models have a surface density distribution that falls
off as $\Sigma\propto R^{-1.5}$ for $R<200$ AU; beyond 200 AU the surface 
density drops off as $\Sigma\propto R^{-12}$ to simulate an effective outer 
edge of 200 AU while keeping a continuous mass distribution. In the B series of
models the total disk mass is varied from $10^{-1}\,M_\odot$ in B1 to 
$10^{-4}\,M_\odot$ in B4, while keeping a constant dust/gas mass ratio of 0.01.
In the BL series the gas mass stays constant at $10^{-1}\,M_\odot$ and the 
dust/gas ratio is varied from 0.01 in BL1 to $10^{-5}$ in BL4. \citet{duldom04}
kept the overall dust/gas ratio in the BL series constant at 0.01 by removing 
the interstellar-sized grains ($\sim 0.1\,\mu m$) from the disk and replacing 
them with cm-sized 
grains in the midplane, thus simulating the effects of dust settling and 
growth, but these large grains are excluded from the 
chemical calculations presented here because of their very limited influence
\citep[e.g.][]{kamzad01,jonkhe06}. It should be noted that models B1 and BL1 
are identical. The abundance of polycyclic aromatic 
hydrocarbons (PAHs) is assumed to follow the dust/gas ratio (i.e. the PAH 
abundance drops off with a factor 10 in each step of the BL series), consistent 
with the findings of \citet{ackanc04} and \citet{habart04} that PAH emission is
weaker in group II sources than in group I sources. The abundances of the 
elements used in the calculations, given in Table \ref{elabun}, 
 are representative for disks where N and O are mostly depleted in some 
form (e.g. ${\rm H_2O}$ ice) onto dust rains.

The densities and dust temperatures are taken directly from the 
\citet{duldom04} models. Only regions with densities 
$n_{\rm H}>10^3\,{\rm cm^{-3}}$ were taken into account, regions with lower
densities do not contribute enough column density to be taken into 
consideration. Note that the dust temperature does not fall below
30 K anywhere in the disk, so the freeze-out of CO can be disregarded. The 
maximum outer radius of the computational grid is taken to be 500 AU.

In the B series (Figure \ref{dd04b}) the disk is initially very large and 
flared. As the mass decreases in this series, the disk becomes smaller in both 
the radial and the vertical direction. Although the disk is still rather thick
in models B3 and B4, the $\tau=1$ surface for stellar radiation shows that 
there is little flaring: the disk does not intercept much radiation at large 
radii. These models are a transition between group I sources (of which B1 and 
B2 are examples) and group II sources (B5 and B6), which have too low gas 
masses to be considered in this paper. The structure of the dust temperature 
and the 
radiation field show that the inner part of the disk is strongly self-shadowed.
As the disk mass decreases, this effect becomes less strong and the low dust 
temperatures and UV intensities are confined to a narrow region near the 
midplane.

The BL series is very similar to the B series regarding the temperature 
structure and UV intensities. While one would expect
more flaring in models BL3 and BL4 (since the gas distribution is higher in the
outer disk), the $\tau=1$ surface again shows that these disks have a largely
flat geometry. As in the B series, these models form a transition between group
I (BL1 and BL2) and group II (BL5 and BL6) sources.

\begin{figure*}[!tp]
\centering
\includegraphics[width=17cm]{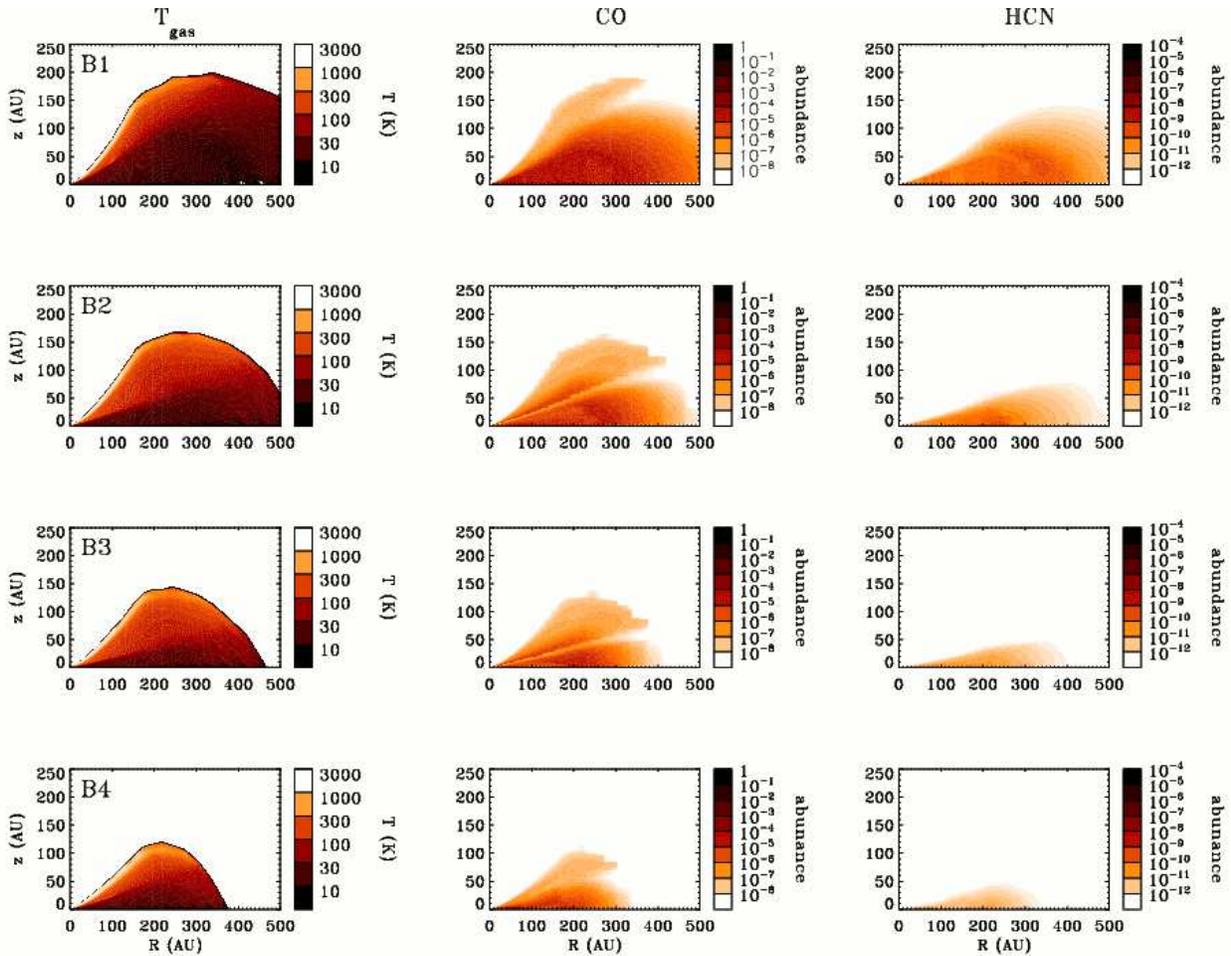}
\caption{Gas temperatures (left column), CO abundances (middle column) and HCN abundances (right column) for models B1--B4.}
\label{2dchemb}
\end{figure*}

\begin{figure*}[!tp]
\centering
\includegraphics[width=17cm]{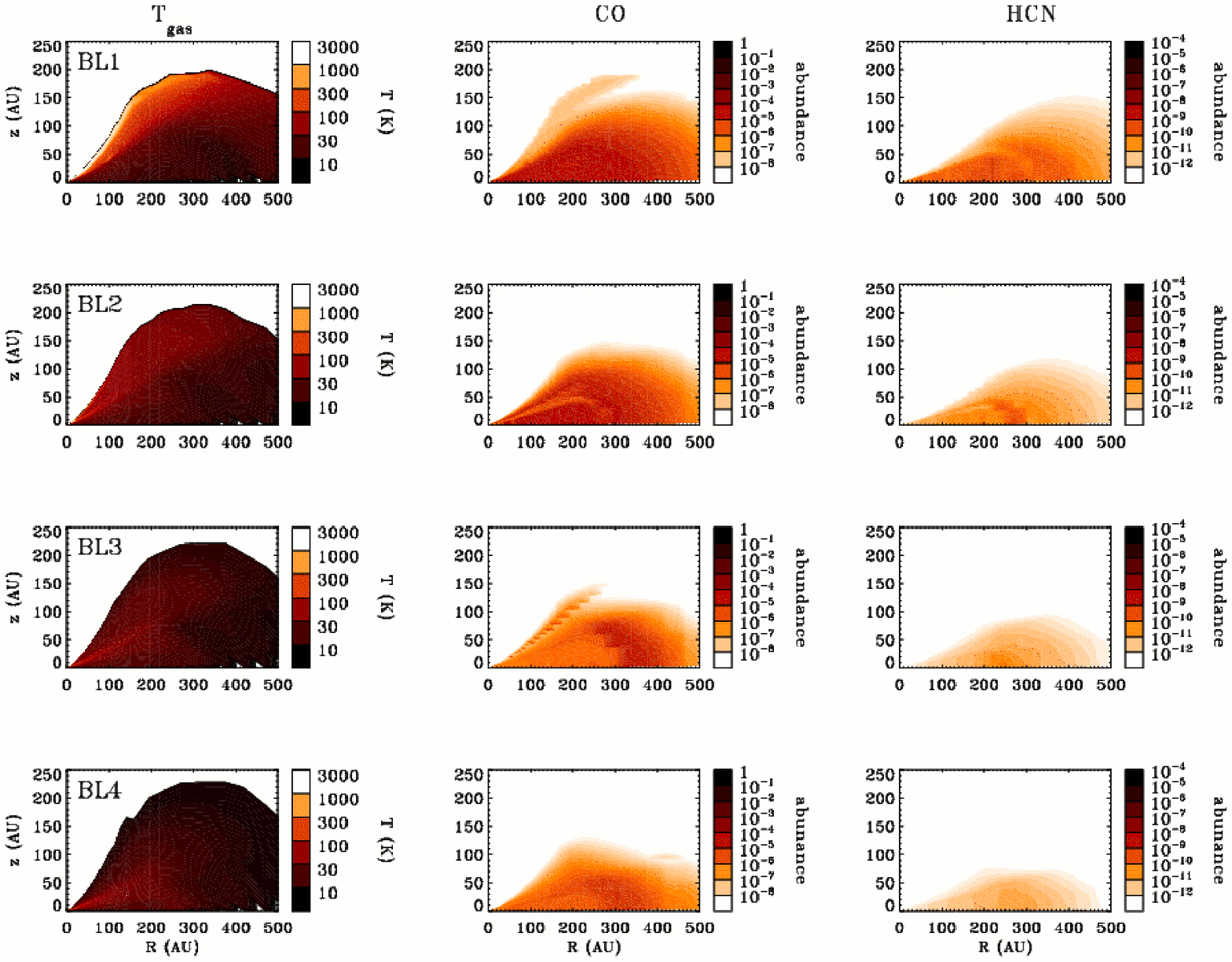}
\caption{Gas temperatures (left column), CO abundances (middle column) and HCN abundances (right column) for models BL1--BL4.}
\label{2dchembl}
\end{figure*}

The transfer of UV radiation is performed with the 2-D Monte Carlo code by 
\citet{zadelh03} to account
for the effects of forward scattering and to include the interstellar 
radiation field. The self-shielding of ${\rm H_2}$ and CO
are included by assuming constant abundances for these species of 0.5 and 
$10^{-5}$, respectively; the results do not depend strongly on the values 
used. The shielding factors are calculated with the mean 
column density computed by the Monte Carlo code 
\citep[for more details, see][]{zadelh03}. The results of the UV transfer are
used to calculate the photorates directly from the cross-sections and local 
radiation field \citep[following][]{dishoe88} and
photoelectric heating rates on classical grains and PAHs 
\citep[][respectively]{tiehol85,baktie94}.

For the stellar radiation field a NEXTGEN spectrum was adopted with 
$T_{\rm eff}=10\,000$ K, $\log g =4$ 
\citep[][see Figure \ref{starspec}]{hausch99} and a stellar radius 
of $2 R_\odot$. The interstellar radiation field was taken from 
\citet{draine78}, with the extension for 
$\lambda > 2000$ \AA\ by \citet{disbla82}. Since the formulation of the
photoelectric heating rates does not take the detailed spectrum of 
the local radiation field into account, the correction factor by 
\citet{spaans94} (Equation 13 in that paper) was used to account for the 
$10\,000$ K stellar spectrum. The value $I_{\rm UV}$ characterizing the 
strength of the UV field was obtained by integrating the radiation field 
between 912 and 2050 \AA\ and dividing this by the same integral for the 
\citet{draine78} field.

With the photorates and photoelectric heating rates known, the chemistry and 
gas temperature are calculated as outlined in \citet{jonkhe04}: the chemical
network of \citet{jansen95} is used to calculate the gas-phase chemistry in 
steady state, and the thermal balance is solved using an escape probability 
formalism assuming cooling radiation escapes from the disk in the vertical 
direction (i.e., column densities used in the escape probability are calculated
vertically). The heating processes included
in the calculations are: photoelectric effect on both classical grains and 
PAHs, cosmic ray ionization of H and ${\rm H_2}$, ${\rm H_2}$ formation,
${\rm H_2}$ photodissociation, collisional de-excitation of the excited 
vibrational levels of ${\rm H_2}$ and the fine structure levels of O, and C 
photoionization. The gas is cooled by line emission from the fine structure 
lines of ${\rm C^+}$, C and O, the rotational lines of CO, and thermal coupling
with dust grains. A more thorough description of these processes can be found 
in \citet{jonkhe04}.

The line profiles are calculated from the chemistry and gas temperature using
the 2-D cylindrical symmetric accelerated Monte Carlo code RATRAN by 
\citet{hogtak00}. To speed up 
the calculations only the material outside $R=10\,{\rm AU}$ is considered, 
since the inner part of the disk contributes little to the overall emission of 
the atomic and molecular lines under consideration in a typical submm observing
beam ($\sim 15''$). The number of 
grid cells varies between 900 and 1500, depending on the model.

\section{Results and discussion}

\subsection{Gas temperature}

\begin{figure*}[!tp]
\centering
\includegraphics[width=17cm]{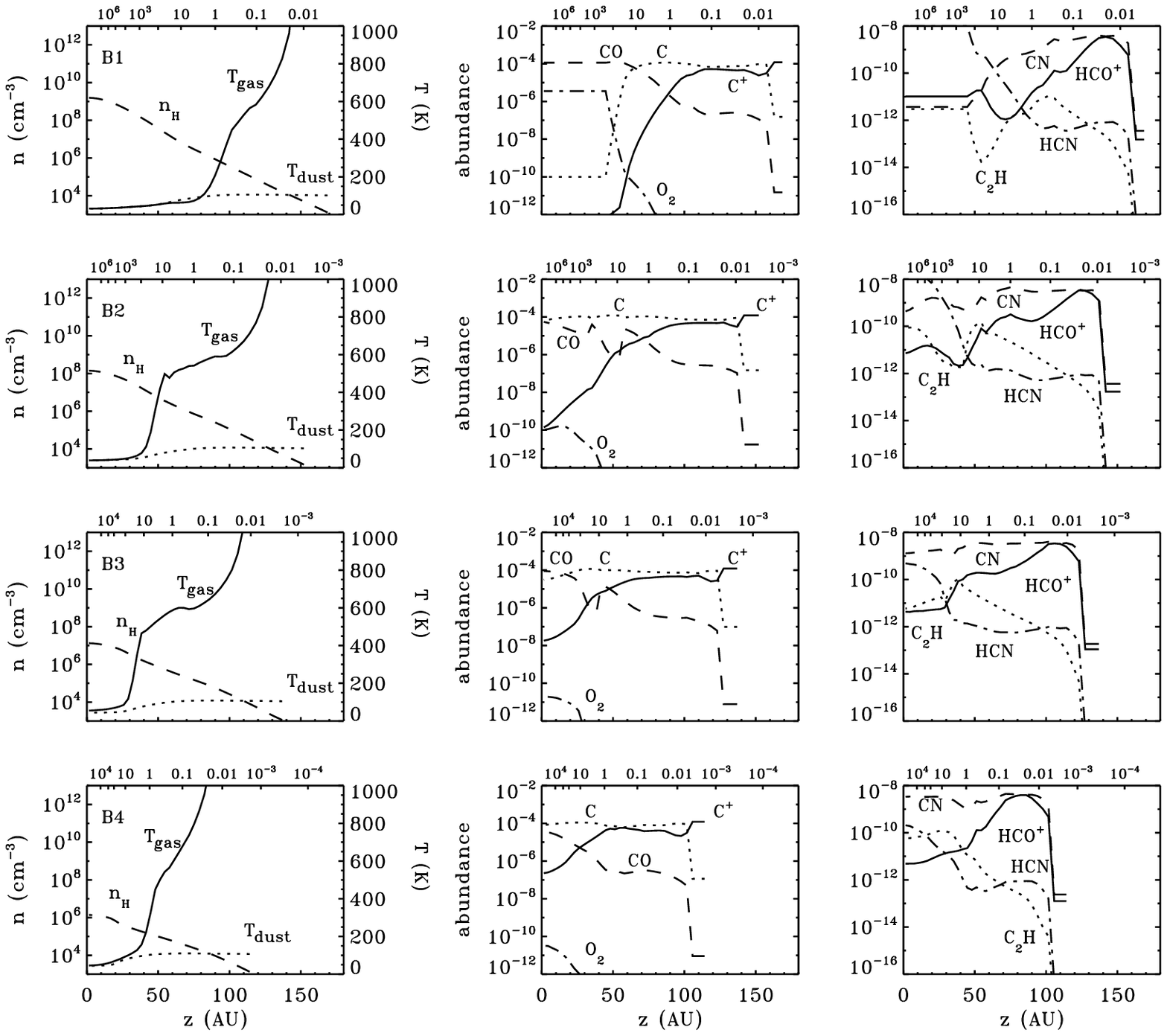}
\caption{Vertical profiles at a radius of 200 AU of the gas temperature and density (left column) and chemistry (middle and right columns) for models B1--B4. The visual extinction toward the central star is given at the top of each panel.}
\label{1dchemb}
\end{figure*}

\begin{figure*}[!tp]
\centering
\includegraphics[width=17cm]{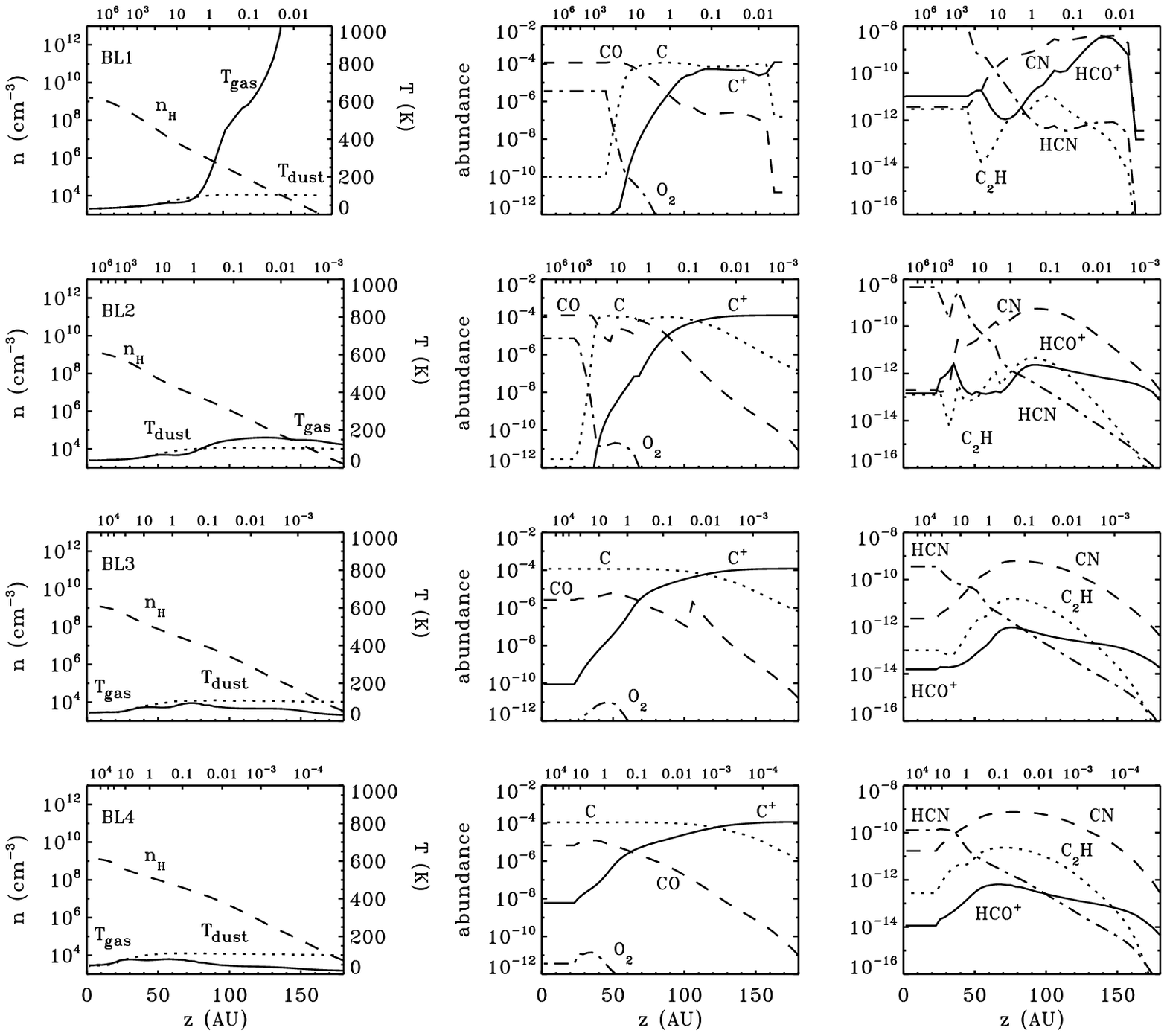}
\caption{As Figure \ref{1dchemb}, but for models BL1--BL4.}
\label{1dchembl}
\end{figure*}

\begin{figure*}[!tp]
\centering
\includegraphics[width=17cm]{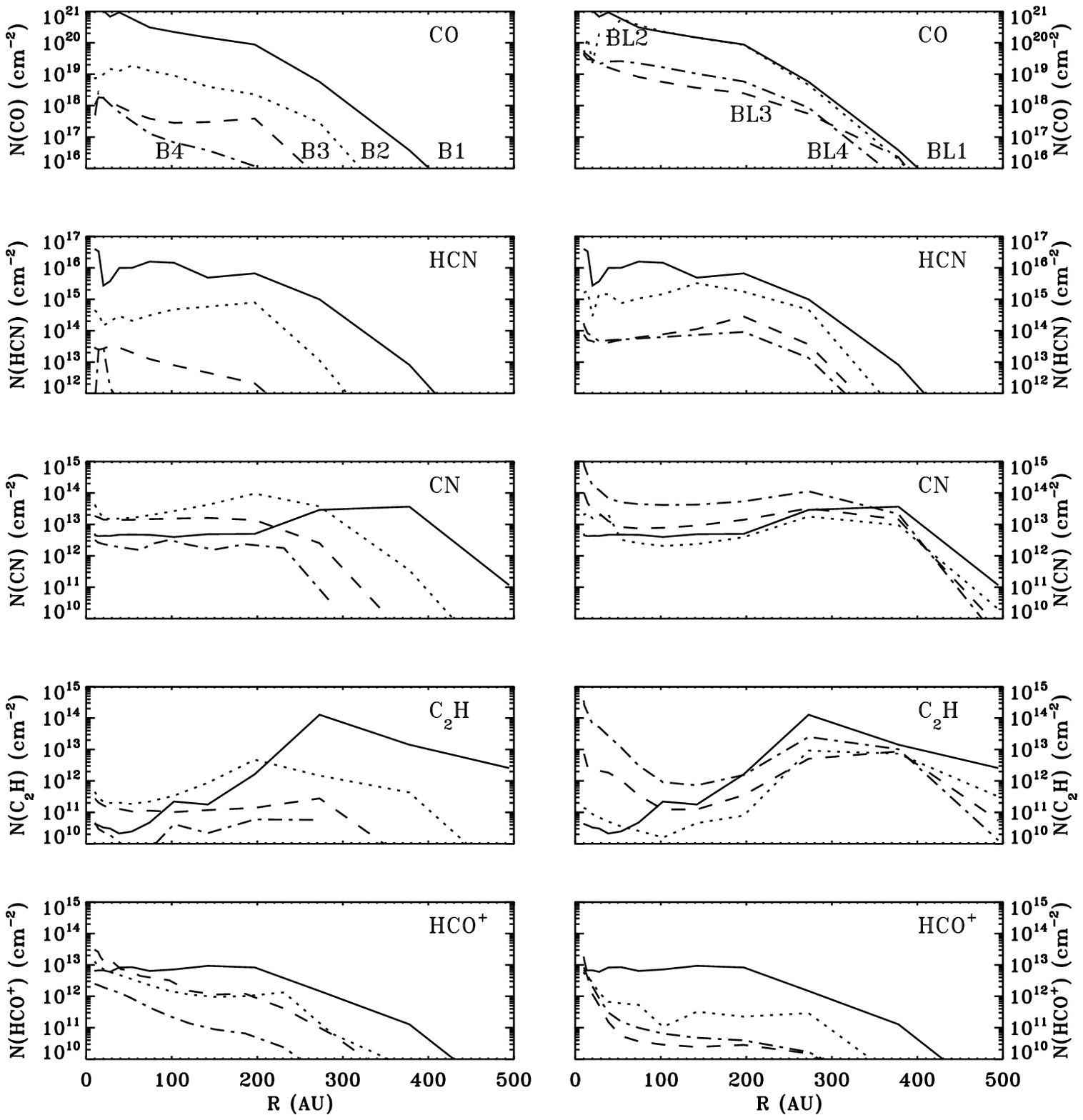}
\caption{The vertical column densities of CO (top row), HCN (second row), CN (third row), ${\rm C_2H}$ (fourth row) and ${\rm HCO^+}$ (bottom row) for the B series (with a descending disk mass, left column) and the BL series (with a descending dust/gas ratio, right column). The solid lines give the results for model B1/BL1; the dotted line for B2/BL2; the dashed line for B3/BL3; the dash-dotted line for B4/BL4.}
\label{columns}
\end{figure*}

The calculated gas temperatures are illustrated in Figures \ref{2dchemb} and 
\ref{1dchemb} for models B1--B4, and in Figures \ref{2dchembl} and 
\ref{1dchembl} for models BL1--BL4. In the B series, the surface of the disk is
very hot (several thousend K) due to the photoelectric heating, which is very 
efficient compared to the cooling by [\ion{O}{i}] and Ly$\alpha$. The 
temperatures found here are consistent with PDR models with 
$n_{\rm H}=10^3\,{\rm cm^{-3}}$ and $I_{\rm UV}=10^4-10^5$. The higher 
densities in 
the optically thin part of the disk below the immediate surface increase the 
efficiency of [\ion{O}{i}] cooling, resulting in lower temperatures (several 
hundred K). Deeper into the disk the densities are so high 
($10^7-10^8\,{\rm cm^{-3}}$) that gas-dust coupling dominates the thermal 
balance and the gas temperature is equal to the dust temperature. In the outer 
parts of the disk ($R>200$ AU) the densities are not high enough for gas-dust
coupling to dominate due to the rapid decline of the total column 
density. In those regions, the gas temperature drops below the dust temperature
due to efficient CO cooling.

It should be noted that the gas 
temperatures found here are higher than those reported in \citet{jonkhe04} for
T-Tauri disks. This is because the radiation fields used in the current paper 
are almost two orders of magnitude larger (see Figure \ref{dd04b}): 
at a radius of 100 AU the current models have $I_{\rm UV}=5-6\times 10^4$ at 
the disk's surface, in contrast with a value of $5-6\times 10^2$ in 
\citet{jonkhe04}. This is due to the more luminous central star used by
\citet{duldom04} and this work.

In the BL series the importance of photoelectric heating decreases with each 
step because both small dust grains and PAHs disappear from the disk. This
means that the total heating rate is less than in the B series, and 
consequently the temperatures at
the disk's surface are lower (20-200 K going from BL4 to BL2). In models BL3
and BL4 heating by C ionization becomes important, and this heating rate has
a similar magnitude as the photoelectric processes. At large depths, gas/dust 
coupling dominates the thermal balance, even with the low dust/gas ratio in 
these models. Since the gas temperature is lower than the dust temperature in 
much of the disk, this means that the temperature increases toward the 
midplane.
  
\subsection{Chemistry}

The vertical distribution of chemical species resembles the structure commonly
found in PDRs: a thin layer of mostly atomic hydrogen at the surface, while
slightly deeper in the disk ${\rm H_2}$ becomes self-shielding and thus 
hydrogen becomes mainly molecular. At the surface carbon is mostly in the form 
of ${\rm C^+}$ and converts to C and CO deeper in the disk. At very high 
optical depths atomic oxygen is locked into ${\rm O_2}$. 

There are several
deviations from the classical PDR structure. In the B series there is a layer
of relatively high CO abundance in the optically thin part of the disk (at 
$z\gtrsim 50$ AU), where
CO is not self-shielding. This is caused by the high temperatures in this 
layer, which help formation of CO via the reaction ${\rm C^++H_2\to CH^++H}$, 
which has an enery barrier of 4600 K. ${\rm CH^+}$ in turn reacts with O to 
form ${\rm CO^+}$, which leads to ${\rm HCO^+}$ and CO. Although CO is
completely self-shielded in the midplane of most of the models (an exception
being the outer regions in model B4, where the CO photodissociation by the 
interstellar UV field is unshielded), the abundance of C is still relatively 
high there. This is caused by the shape of the stellar radiation field (see
Figure \ref{starspec}): there are still many continuum photons with
$\lambda>1200$ \AA\ which can dissociate precursor molecules like CH, which is
an important pathway of CO formation. In the later steps of the BL series C is
the dominant form of carbon in the midplane instead of CO. This is partly due
to the photodissociation of precursor molecules, and partly because the 
formation of CO is further suppressed due to lack of PAHs in these models. The
reaction ${\rm PAH:H + C^+\to PAH+CH^+}$ (where PAH:H is a PAH with an extra H 
atom) is another important step in the reaction pathway leading to CO 
\citep[for rate coefficient, see][]{canosa95}. In the 
later steps of the BL series the PAH abundance is very low, and so this pathway
is closed.

The global distributions of CO and HCN are illustrated in Figure \ref{2dchemb} 
for models B1--B4, and in Figure \ref{2dchembl} for models BL1--BL4. The 
abundances
of CO and HCN decrease significantly going from model B1 to B4, while this 
effect is less pronounced going from BL1 to BL4. Although the UV continuum flux
is comparable in both series, the total gas mass stays at 0.1 $M_\odot$ in the 
BL series while it decreases with factors of 10 in the B series. The 
self-shielding of CO will therefore be much stronger in the BL series, 
resulting in higher CO abundances. 
HCN is rapidly dissociated in the optically thin regions of both series,
but its formation rate correlates with CO abundance. Therefore, a rapid 
decrease in HCN abundance can be seen in the B series, where there is not 
enough CO to counter the rapid photodissociation. In the BL series, however,
the high CO abundances increase the production rate of HCN, keeping it at a 
higher abundance than in the B series.

A more detailed view of the chemistry is given in Figures \ref{1dchemb} and 
\ref{1dchembl}, which show a vertical cut at 200 AU. In the B series the high 
CO abundances at the surface result in high abundances of ${\rm HCO^+}$, 
${\rm C_2H}$, CN and HCN. 
The abundance of ${\rm C_2H}$ follows the same trends as CN, which
is in agreement with the results by \citet{zadelh03} for T-Tauri stars. Note 
that in contrast
with \citet{zadelh03} (who found similar abundances for these molecules) the 
abundance of ${\rm C_2H}$ found here is one to two 
orders of magnitude less than CN. This is due to the difference in stellar 
spectrum used in the current paper, which has a jump in intensity at 
wavelengths longer than 1200 \AA. Since ${\rm C_2H}$ can be dissociated by 
photons with $\lambda>1200$ \AA\ and CN is dissociated only at $\lambda<1150$ 
\AA, the photodissociation rate of ${\rm C_2H}$ is much higher, and thus this 
molecule has a lower abundance. 

The same effect is visible in the abundance of
HCN compared to CN: HCN is easily dissociated by photons with $\lambda>1200$
\AA,
giving it a lower abundance than CN in most of the disk. Only in the heavily
shadowed regions near the midplane is HCN more abundant than CN. In the BL
series the densities near the midplane are high enough to increase the HCN
formation rate, and so the HCN abundance is higher than CN even in the 
optically thin models BL3 and BL4.

The abundances of $\rm HCO^+$ in the deeper layers of the disk are lower by 
several orders of magnitude than those found by \citet{zadelh03}. A possible 
explanation for this may be that charge exchange with metals plays a more 
important role in the current model, since the densities are a factor 
$\sim 100$ higher. Also, as for CO, the rapid photodissociation of hydrocarbon 
precursors plays a role.

In summary, the B series shows a clear trend in molecular abundances: going 
from model B1 
to B4 molecules are more easily dissociated, and the peaks in their vertical 
distributions occur closer to the midplane. In the BL series the trend is less 
clear. Going from model BL1 to BL2, the drop in gas temperature greatly affects
the carbon chemistry. The differences between models BL2 to BL3 are limited, 
and models BL3 and BL4 are nearly identical. Again there is a shift 
toward the midplane in the peak of the vertical abundance distribution, in 
agreement with earlier findings of \citet{jonkhe04} and \citet{aiknom06}, but 
this shift is small compared to the B series.


\subsubsection{Influence of PAHs} 
It is assumed that PAHs are present at interstellar abundances 
($n_{\rm PAH}/n_{\rm H}=10^{-7}$) for a dust/gas ratio of $10^{-2}$, and that 
these abundances scale with the dust/gas ratio. This assumption is consistent 
with the findings of \citet{ackanc04} and \citet{habart04}, although models
by Geers et al. (in press) suggest an abundance that is lower by as much as an 
order of magnitude.
The influence of PAHs on the chemistry can best be seen in the BL 
series (Figure \ref{1dchembl}), where the PAH abundance is decreased by a 
factor 10 at each step. PAHs affect the chemistry mainly through charge 
exchange reactions (of the form ${\rm PAH + X^+\to PAH^+ + X}$) and 
hydrogenation reactions (of the form ${\rm PAH:H + X \to PAH+XH}$). 
Especially the latter reactions are important in the formation of molecules, as
can be seen in the distribution of CO in the BL model series (seen in 
Figures \ref{2dchembl} and \ref{1dchembl}), where the PAH abundance is 
decreased in each step. As a result, the reaction 
${\rm C^+ + PAH:H\to CH^+ + PAH}$ becomes less efficient; since ${\rm CH^+}$ is
a chemical precursor of CO (via ${\rm CO^+}$ and ${\rm HCO^+}$), the overall CO
abundance decreases together with the PAH abundance.

\subsubsection{Carbon chemistry}
In some of the models, particularly BL3 and BL4, there is a broad layer near the
disk midplane where atomic carbon is the main carbon-bearing species. This 
somewhat unconventional result is caused by shape of the stellar radiation 
field, shown in Figure \ref{starspec}: the stellar radiation has few photons in
the 912-1200 \AA\ range, most of the flux is at longer wavelengths. This
means that the photoionization rate of C and the photodissociation rate of CO 
are small, since these processes occur in the 912-1100 \AA range. The 
photodissociation rate of CH is relatively high, however, since this molecule 
is dissociated at longer wavelengths than CO. Since CH is an important chemical
precursor to CO, the rapid dissociation of CH means that the formation rate of 
CO is very low, resulting in a low CO abundance. Since atomic carbon is 
released by the photodissociation of most carbon-bearing molecules, and the 
ionization rate of atomic carbon is very low, carbon is mostly neutral and 
atomic.

\subsection{Column densities}

The vertical column densities of CO, CN, HCN, ${\rm C_2H}$ and ${\rm HCO^+}$
are shown in Figure \ref{columns}. It can be seen that the greatest variations
occur in the B series for CO and HCN. The CO abundance is sensitive to the gas 
mass due to self-shielding. Similarly, HCN occurs only in regions that are 
dense and have sufficient continuum optical depth toward the central star. Both
of these
conditions are strongly dependent on the disk mass. The column densities of 
CN and ${\rm C_2H}$ in the B series show a peak which starts at relatively 
large radii and shifts inward going from B1 to B4. These molecules require a 
moderate optical depth toward the central star: large enough so they and their 
precursors are not dissociated, but small enough that they can still be formed 
by photodissociation of less stable molecules (HCN for CN; ${\rm C_2H_2}$ and
${\rm C_3H}$ for ${\rm C_2H}$). The ${\rm HCO^+}$ peak also shifts inward due
to the lower abundance of CO in the outer regions of the disk.

\begin{figure*}[!tp]
\centering
\includegraphics[width=17cm]{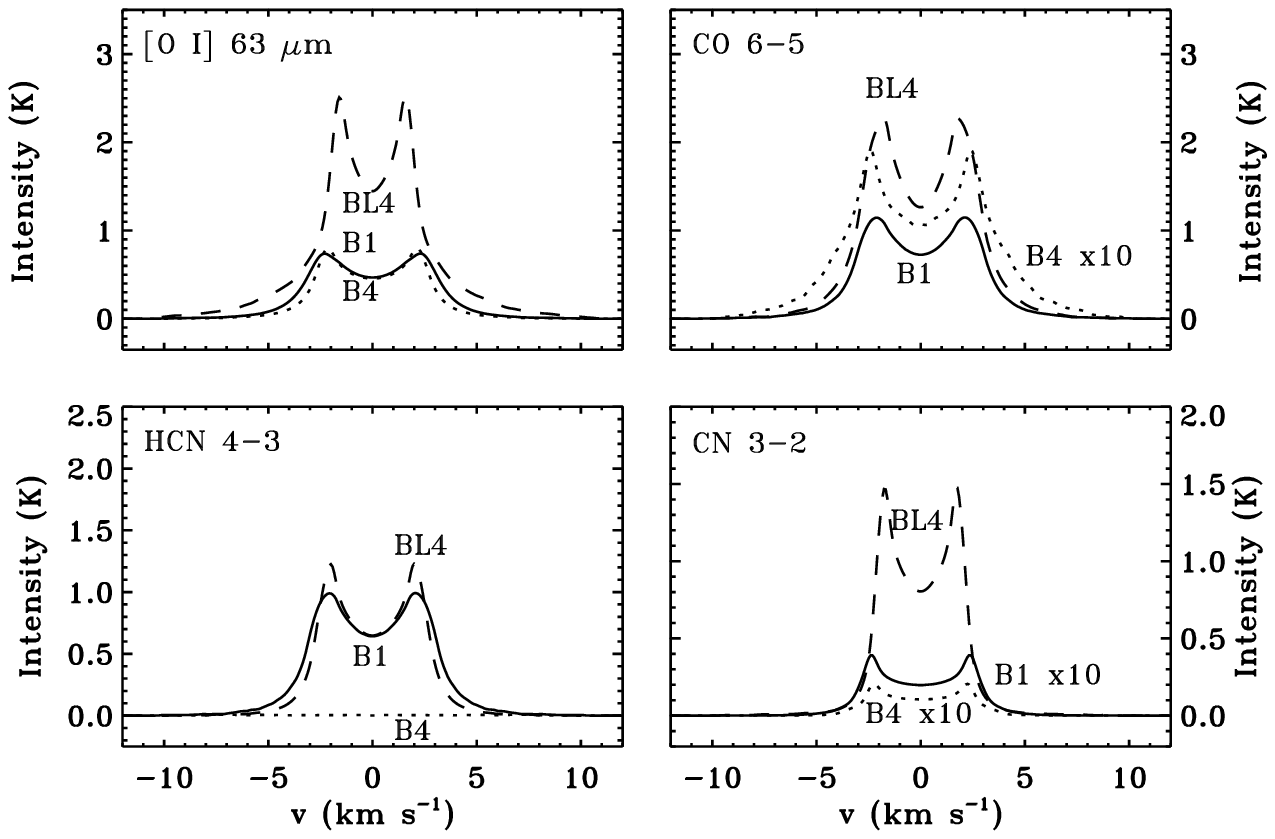}
\caption{Emission line profiles for O (top left), CO (top right), HCN (lower left) and CN (lower right) for models B1/BL1 (solid line), B4 (dotted line) and BL4 (dashed line).}
\label{lines}
\end{figure*}

In the BL series, the CO column densities show a clear distinction between on 
the one side models BL1 and BL2, and on the other side BL3 and BL4. The reason 
for this can be seen in Figure \ref{1dchembl}: in models BL1 and BL2, CO is the
main form of carbon in the dense regions near the midplane. In BL3 and BL4, C 
becomes
the dominant form of carbon due to the low PAH abundance in these models. In 
the BL series HCN occurs mainly in the midplane of the disk, in a layer that 
gets narrower going from BL1 to BL4. This explains the decreasing column 
densities in Figure \ref{columns}. The peaks in the vertical distributions of
CN and ${\rm C_2H}$ get broader going from BL1 to BL4, and consequently 
their column densities in the main portion of the disk ($R<200$ AU) increase. 
In the outer disk the
interstellar radiation field can dissociate these molecules, decreasing their 
column densities. The peak in the vertical ${\rm HCO^+}$ distribution decreases
in height going from model BL1 to BL4, which is reflected in the column 
densities.

The column densities of CN and ${\rm C_2H}$ (and to a 
lesser extent those of ${\rm HCO^+}$) are relatively high in the outer parts of
the disk ($R>200$ AU). In spite of the fact that the total column density drops
off as $R^{-12}$, the densities are still high enough to provide high 
abundances for these molecules. The abundances of these molecules in the outer 
disk depend strongly on the exact form of the total column density profile. If 
it drops off more rapidly than $R^{-12}$, this will be reflected in 
the intensities of the molecular lines. As a consequence, the emission 
lines of CN, ${\rm C_2H}$ and ${\rm HCO^+}$ may be used to examine the outer 
regions of disks that are difficult to probe in the continuum.

The trends in the column densities of CO, CN and ${\rm HCO^+}$ presented here 
for the BL series, in which dust growth and settling is simulated, show 
qualitative similarities to those found by \citet{aiknom06} for T-Tauri 
disks with different stages of grain growth. With increasing grain size, they 
find that $N({\rm CO})$ changes only very little, $N({\rm CN})$ stays constant
in the outer regions and increases in the inner regions, and $N({\rm HCO^+})$ 
decreases in the outer regions. They find no significant variation
for HCN and ${\rm C_2H}$, whereas column densities of these molecules change
significantly in the current paper. This is possibly due to the jump in the 
stellar radiation field at 1200 \AA, which has a strong influence on these 
molecules; \citet{aiknom06} use a much
smoother stellar spectrum that resembles the interstellar radiation field.

\subsection{Atomic and molecular lines}

\begin{figure*}[!tp]
\centering
\includegraphics[width=17cm]{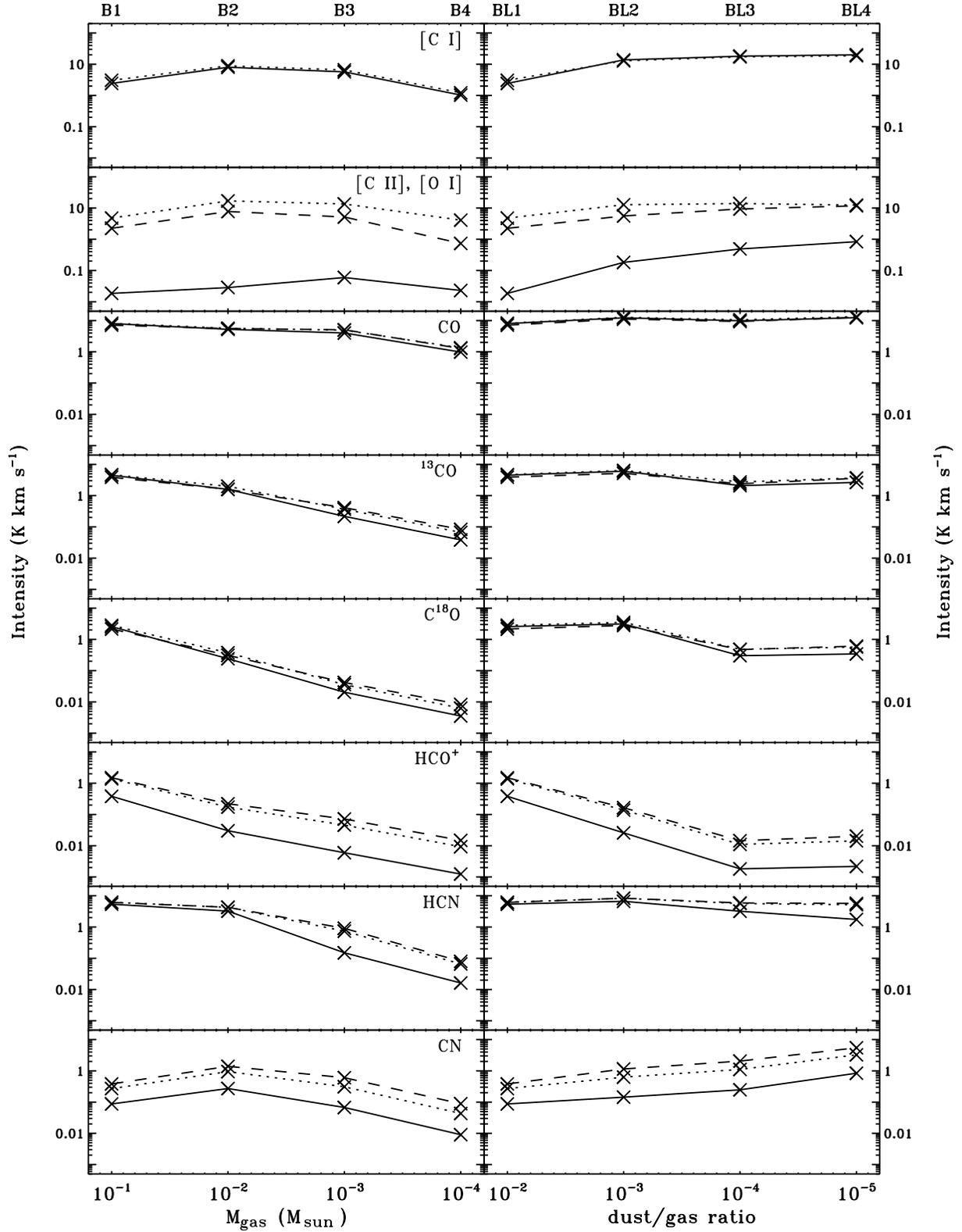}
\caption{Integrated intensities in a 6.7'' beam for the B series (left column) and the BL series (right column). The top row gives the [\ion{C}{i}] fine structure lines ($J=1-0$ solid, $J=2-1$ dotted). The second row gives the fine structure lines of [\ion{C}{ii}] fine structure line (solid) and [\ion{O}{i}] ($J=1-2$ dotted, $J=0-1$ dashed). The third, fourth and fifth rows give the rotational lines of CO, ${\rm ^{13}CO}$ and ${\rm C^{18}O}$ ($J=2-1$ solid, $J=3-2$ dotted, $J=6-5$ dashed). The sixth row gives the rotational lines of ${\rm HCO^+}$ ($J=1-0$ solid, $J=3-2$ dotted, $J=4-3$ dashed). The seventh row gives the rotational lines of HCN ($J=1-0$ solid, $J=3-2$ dotted, $J=4-3$ dashed). The bottom row gives the rotational lines of CN ($J=1-0$ solid, $J=2-1$ dotted, $J=3-2$ dashed).}
\label{ints}
\end{figure*}


From the molecular abundances and the gas temperature at each position in the 
disk the profiles of rotational molecular lines are calculated using the 
accelerated Monte Carlo code by \citet{hogtak00}. The distance to the disk is 
taken to be 150 pc, and an inclination of $45^\circ$ is 
assumed. For the velocity field Keplerian rotation around a $2.5\,M_\odot$
star 
and a turbulent velocity of ${\rm 0.2\,km\,s^{-1}}$ were assumed. The
data were convolved with a 6.7'' beam, so that the disk fills the
beam precisely. For molecules containing ${\rm ^{13}C}$ and 
${\rm ^{18}O}$ the abundances of the main isotopic variant were used, scaled 
with factors of 1/45 and 1/500, respectively. This assumption does not hold 
everywhere in the disk due to isotope selective photodissociation, but the 
effects on integrated intensities over the entire disk are likely small.
The resulting emission line profiles for O, CO, HCN 
and CN for models B1/BL1, B4 and BL4 are shown in Figure \ref{lines}. 

It can be seen that the [\ion{O}{i}] line does not change much when the mass of
the disk decreases; the difference in intensity between models B1 and B4 is 
only a factor of a few, while the mass changes with a factor of 1000. Model BL4
has a slightly higher [\ion{O}{i}] intensity, since at the temperatures where O
can be excited, less oxygen is in the form of CO and ${\rm O_2}$. Also, the 
[\ion{O}{i}] line has broader wings since it is excited mostly in the inner 
regions of the disk compared to model B1.

The CO $J=6-5$ line follows similar trends: it is optically thick in all 
models, making it relatively insensitive to total disk mass. The line is
significantly broader in model B4 than in models B1 and BL4 because the CO
column densities are more centrally concentrated in that model.

The HCN $J=4-3$ line shows a strong dependence on the disk mass. Model B4 has
very low HCN abundances, giving this line an intensity in the mK range. The 
differences between models B1 and BL4 are less pronounced and reflect the 
slightly lower HCN abundances in BL4 with respect to B1. The CN $J=3-2$ line,
however, shows a strong dependence of the line shape on the dust/gas ratio.
This line has a relatively low excitation temperature ($\sim 30$ K), and the 
gas temperature in the region where it can be excited (i.e. where the density 
is equal to the critical density of $10^7\,{\rm cm^{-3}}$) is too high in model
BL4 to get an appreciable population in the upper level. In model B1 the
temperatures are lower due to the decreased C-ionization heating rate, and the
$J=3$ level has a higher population. Model B4 only has a weak line here due to
the low gas content of the disk.

The integrated intensities are presented in Figure \ref{ints}.
It can be seen that model B1/BL1 generally produces strong emission lines,
especially for the fine-structure transitions of O and the rotational lines of 
CO. These lines are the main coolants of the PDR surface of the disk; therefore
they radiate all the energy that is delivered to the gas via the various 
heating processes. 

The [\ion{C}{i}] intensities show an initial increase in the B series even 
though the disk mass decreases by a factor 10, this is due to the high C
abundance in model B2 compared to B1. In models B3 and B4 the intensity drops 
again, albeit slowly due to the optical thickness of the lines. In the BL 
series the [\ion{C}{i}] intensities only increase due to the ever increasing C
abundances in this series. Similar effects also determine the intensities of 
the [\ion{O}{i}] line.

\begin{figure}
\resizebox{\hsize}{!}{\includegraphics[angle=0]{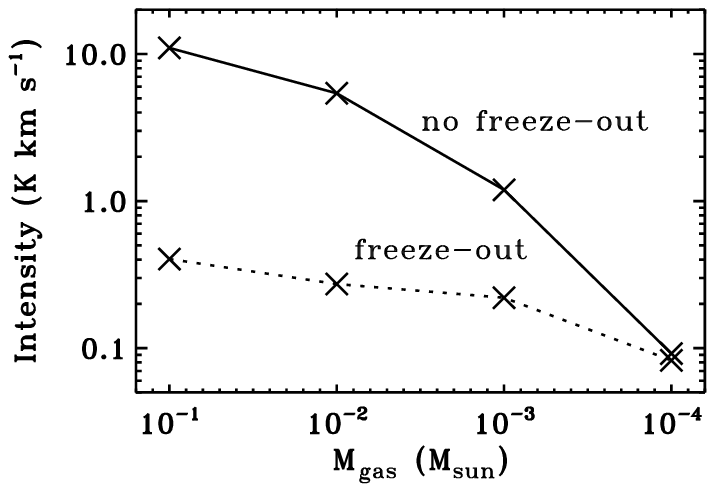}}
\caption{Integrated intensities of HCN in the B series when all HCN is in the gas phase (solid line) and when HCN is removed from the gas phase for $T_{\rm dust}< 70$ K (dotted line). The BL series follows similar trends.}
\label{hcnfr}
\end{figure}

The [\ion{C}{ii}] fine-structure line is significantly weaker than those of 
[\ion{O}{i}] and [\ion{C}{i}], even though it is an important cooling line in 
PDRs.
This line has a low critical density ($10^3\,{\rm cm^{-3}}$), which means 
is can only be an effective coolant at the immediate surface of the disk. 
Deeper into the disk this line is quenched by collisions. Its intensity 
increases in the BL series because the diffuse upper layers are thicker there.

The intensities of the CO rotational lines show a weak dependence on disk mass.
Initially these lines are optically thick, but become optically thin going from
model B3 to B4, causing a sudden drop in intensity. The lines do not change 
significantly in the BL series.

The pure rotational lines of ${\rm HCO^+}$, HCN and CN follow the same trends 
as CO: they change only little in the first steps of the B series, but become 
optically thin going from model B2 to B3 due to their lower abundance compared 
to CO. These lines change very little in the BL series, with the exception of 
${\rm HCO^+}$, which has a steadily decreasing column density in this series.

\subsection{Comparison with observations}

The rotational lines of CO, ${\rm HCO^+}$, CN and HCN of two Herbig Ae 
stars, HD 163296 and MWC 480, have been observed by \citet{thi04} with IRAM and
JCMT. The SEDs of these objects indicate that they are group II sources, 
meaning that of the models presented here only steps 3 or 4 of each series need
to be considered. Correcting for beam dilution, their intensities of CO, 
${\rm ^{13}CO}$
and CN, as well as the CO/${\rm ^{13}CO}$ line ratio, fit well to the values 
found for model B2 (i.e. a $0.01\,M_\odot$ disk with a dust/gas ratio of 0.01)
with the HD 163296 data indicating a slightly higher gas mass than MWC 480.
A good fit is also given by models BL3 and BL4 (i.e. a $0.1\,M_\odot$ disk with
a dust/gas ratio of $10^{-4}-10^{-5}$),
with the HD 163296 data indicating a slightly lower dust/gas ratio than MWC 
480. The masses agree within a factor of a few with those derived from 
millimeter continuum 
observations assuming a dust/gas ratio of 0.01. None of the models presented 
here can reproduce the ${\rm HCO^+}$ lines by \citet{thi04}, which have 
intensities of ${\rm >1\,K\,km\,s^{-1}}$. This may be due to the low 
${\rm HCO^+}$ abundances found here compared to other disk chemistry models.
Since these objects are group II sources, it is unlikely that model B2 
describes their structure. Instead, models BL3 and BL4 provide a good fit to 
the observational data.

The intensities predicted here for the HCN 4--3 line of models  B2, BL3 and BL4
($\sim 5\,{\rm K\,km\,s^{-1}}$) exceed
the upper limits by \citet{thi04} of 0.8 and $0.3\,{\rm K\,km\,s^{-1}}$ 
for HD 163296 and MWC 480, respectively (corrected for beam dilution). 
This discrepancy may be explained by the fact that our models take 
freeze-out of molecules into account only by adjusting the global gas-phase 
elemental abundances, meaning that there is relatively little oxygen (since most of the dust is colder than the water sublimation temperature of $\sim 100$ K) compared to carbon (since all dust is warmer than the CO sublimation temperature of $\sim 20$ K). For nitrogen-bearing species this method is less accurate, however, since molecules that occur mostly in the disk's surface layers (like CN) will not freeze out due to the high dust temperatures there, while molecules that occur mostly in the deeper regions near the midplane (like HCN) are more likely to freeze out on the cold dust grains.
To estimate the effect of freeze-out, the 
intensities were re-calculated by removing the HCN from regions where 
$T_{\rm dust}<70\,{\rm K}$ \citep[consistent with the TPD experiments on 
${\rm H_2S}$ and ${\rm C_2H_2}$ by][]{collin04} . The results are given in 
Figure \ref{hcnfr}. It can be seen that freeze-out may account for more than an
order of magnitude difference in intensity, and that the revised B2 model 
intensity is just consistent with the observed upper limits. The CN lines are 
less affected by freeze-out, since the peak in 
the vertical CN distribution occurs at larger heights than the HCN 
distribution, where the higher dust temperatures and lower densities will 
prevent freeze-out.

\begin{figure*}[!tp]
\centering
\includegraphics[width=17cm]{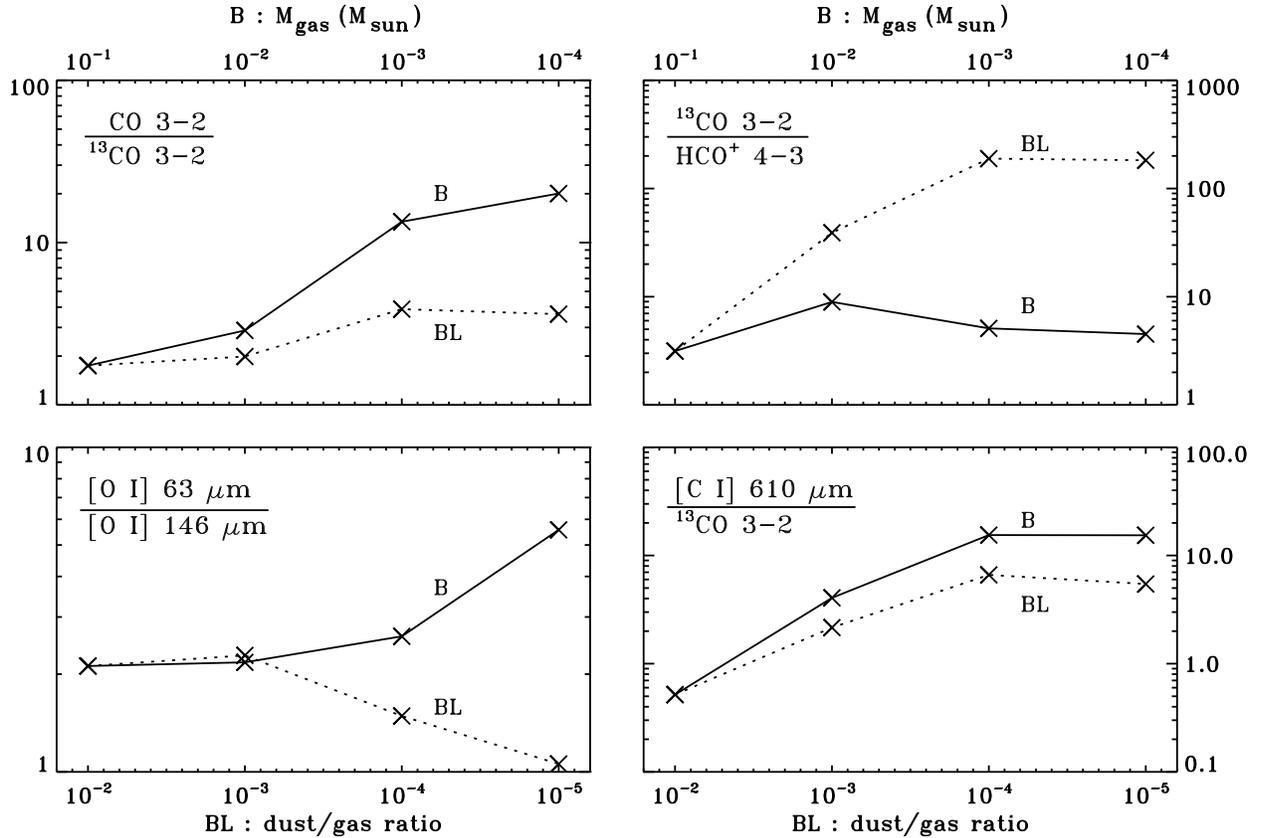}
\caption{Line ratios for the B series (solid lines) and the BL series (dotted lines).}
\label{rats}
\end{figure*}

The intensities presented here indicate that future instruments like ALMA
will be able to detect and image the rotational lines of CO, ${\rm ^{13}CO}$ 
and CN in disks with masses as low as $10^{-4}\,M_\odot$, with a typical 
sensitivity of $0.3\,{\rm K\, km\,s^{-1}}$ in 1 hr. The lines of
${\rm C^{18}O}$, ${\rm HCO^+}$ and HCN can also be imaged in these low-mass 
disks, but will require long integration times. Massive disks that undergo dust
growth and settling will be readily observable, since the molecular column 
densities remain high even when the disk becomes more and more optically thin 
in the continuum.

There are several line ratios that can be used to ascertain whether a group II 
source (as determined by its SED) is the product of mass loss (putting it in 
the B series) or of dust growth and settling (putting it in the BL series). 
Some relevant ratios are plotted in Figure \ref{rats}: a low ratio ($<10$) of 
CO/${\rm^{13}CO}$ ($<10$) or CO/${\rm{C^{18}O}}$ would indicate an 
object with a low dust/gas ratio instead of a low overall mass; 
${\rm ^{12}C^{16}O}$ is optically thick (thus making its lines relatively 
insensitive to the disk mass), while the isotopic
variants of CO are optically thin, so they trace the CO abundance directly.
${\rm ^{13}CO}$ 3-2/${\rm HCO^+}$ 4-3 is another suitable tracer, since the 
${\rm HCO^+}$
intensities behave similarly in the B and BL series; on the other hand, the 
intensities of ${\rm ^{13}CO}$ drop off if the disk mass is decreased, while 
they remain high in the case of dust growth/settling. A high ratio ($> 100$) 
would indicate dust growth and 
settling, while a low ratio would indicate mass loss.
The ratio of [\ion{O}{i}] 63 ${\rm \mu m}$/146 ${\rm \mu m}$ directly probes 
the temperature structure of the upper layes, and is therefore suitable to 
distinguish between the B and BL series. A high ratio ($>2$) would indicate 
that a disk with a dust/gas ratio close to 0.01; a lower ratio would indicate 
dust growth or settling.
The ratio of [\ion{C}{i}] 610 ${\rm \mu m}$/${\rm ^{13}CO}$ 3-2 can be used
to distinguish group I sources (B1--B2 and BL1--BL2) from group II sources 
(B3--B4 and BL3--BL4), where a high ratio ($>5$) would indicate a group II 
source. This ratio is not very sensitive to the differences between the B and 
BL series, however. The CO 6-5/2-1 ratio was found to be insensitive to the 
differences between the B and BL series, since the gas temperatures stay high 
enough to excite the $J=6$ even in models with a very low dust/gas ratio (e.g.
model BL4). 

\section{Conclusions}
The thermal balance, chemistry and emission line profiles were calculated for a
series of disk structure models representing different classes of 
protoplanetary disks around Herbig Ae stars with systematically decreasing disk
masses or dust/gas ratios. The following can be concluded:
\begin{itemize}
\item For an interstellar dust/gas ratio (0.01), the gas temperatures 
at the flaring edge of the disk are very high, 
reaching temperatures of several thousand K. Directly underneath this
hot surface layer is a warm (several hundreds of K) optically thin layer 
where the gas is heated directly by the stellar radiation. This warm layer 
drives an endothermic chemistry resulting in a relatively high CO abundance. In
the midplane and in the outer disk the self-shadowing of the disk results in 
lower gas temperatures, usually close to the local dust temperature. This 
relatively cool self-shadowed part decreases in size with increasing disk mass.

\item The disk chemistry shows a strong dependence on disk mass. The low-mass 
disks have large optically thin atmospheres where fragile species like ${\rm C_2H}$ and HCN are easily destroyed by UV radiation. More robust 
species like CO and CN still have relatively high abundances here.

\item Although the self-shielding of CO is stronger in the BL series compared 
to the B series, the abundance of this molecule is lower. This is due to the
reduced efficiency in CO formation in the absence of PAHsand classical grains, 
and the high photodissociation rates of CO precursors like CH.

\item Dust settling has little effect on molecular abundances if the total
gas mass is kept constant: there is only a small shift of the maximum abundance
towards the midplane. The primary effect of dust settling on the disk is to 
lower the gas temperature by decreasing the importance of the photoelectric 
heating process.

\item The CO rotational lines can be used to trace the total disk mass only for
low-mass disks. In high-mass disks these lines are optically thick and not 
sensitive to the disk mass. The isotopic variants of CO (${\rm ^{13}CO}$ and
${\rm C^{18}O}$) are much more sensitive even at high disk masses; their 
intensities are also mildly sensitive to the dust/gas ratio.

\item Good tracers to distinguish between the B series and the BL series (i.e.
whether a class II source is undergoing mass loss or dust growth and settling)
are CO/${\rm ^{13}CO}$ and CO/${\rm C^{18}O}$ line ratios, since the isotopic 
variants of CO are more sensitive to the total amount of CO in the disk than 
the main form. Also, the ${\rm ^{13}CO}$ 3-2/${\rm HCO^+}$ 4-3 and the 
[\ion{O}{i}] 63 ${\rm \mu m}$/146 ${\rm \mu m}$
ratios can provide information on whether a source belongs in the B or the BL 
series. [\ion{C}{i}]/${\rm ^{13}CO}$ can be used to distinguish group I from
group II sources, but is not sensitive to the differences between a disk 
undergoing mass loss or dust growth and settling.

\item In the disk models presented here, the dust temperatures are high enough 
to prevent the freeze-out of CO onto dust grains. For other molecules, like 
HCN, freeze-out remains possible, but has not been included explicitly in the
current code.

\end{itemize}

\begin{acknowledgements}
The authors are grateful to Inga Kamp and Xander Tielens for their helpful 
comments. This work was supported by a Spinoza grant from the Netherlands 
Organisation 
for Scientific Research and by the European Community's Human Potential 
Programme under contract HPRN-CT-2002-00308, PLANETS.
\end{acknowledgements}

\bibliographystyle{aa}
\bibliography{5668.bib}

\begin{thebibliography}{36}
\expandafter\ifx\csname natexlab\endcsname\relax\def\natexlab#1{#1}\fi

\bibitem[{Acke \& {van den Ancker}(2004)}]{ackanc04}
Acke, B. \& {van den Ancker}, M.~E. 2004, \aap, 426, 151

\bibitem[{Aikawa \& Nomura(2006)}]{aiknom06}
Aikawa, Y. \& Nomura, H. 2006, \apj, accepted

\bibitem[{Aikawa {et~al.}(1997)Aikawa, Umebayashi, \& Miyama}]{aikawa97}
Aikawa, Y., Umebayashi, T., \& Miyama, S.~M. 1997, \apj, 486, L51

\bibitem[{Bakes \& Tielens(1994)}]{baktie94}
Bakes, E. L.~O. \& Tielens, A. G. G.~M. 1994, \apj, 427, 822

\bibitem[{{Bary} {et~al.}(2003){Bary}, {Weintraub}, \& {Kastner}}]{bary03}
{Bary}, J.~S., {Weintraub}, D.~A., \& {Kastner}, J.~H. 2003, \apj, 586, 1136

\bibitem[{Brittain {et~al.}(2003)Brittain, Rettig, Simon, Kulesa, DiSanti, \&
  Russo}]{britta03}
Brittain, S.~D., Rettig, T.~W., Simon, T., {et~al.} 2003, \apj, 588, 535

\bibitem[{Canosa {et~al.}(1995)Canosa, Laub\'e, Pasquerault, Gomet, \&
  Rowe}]{canosa95}
Canosa, A., Laub\'e, S., Pasquerault, D., Gomet, J.~C., \& Rowe, B.~R. 1995,
  Chem. Phys. Lett., 245, 407

\bibitem[{Chiang \& Goldreich(1997)}]{chigol97}
Chiang, E.~I. \& Goldreich, P. 1997, \apj, 490, 368

\bibitem[{Collings {et~al.}(2004)Collings, Anderson, Chen, Dever, Viti,
  Williams, \& McCoustra}]{collin04}
Collings, M.~P., Anderson, M.~A., Chen, R., {et~al.} 2004, MNRAS, 354, 1133

\bibitem[{{D'Alessio} {et~al.}(1998){D'Alessio}, {Cant\'o}, {Calvet}, \&
  {Lizano}}]{daless98}
{D'Alessio}, P., {Cant\'o}, J., {Calvet}, N., \& {Lizano}, S. 1998, \apj, 500,
  411

\bibitem[{Dent {et~al.}(2005)Dent, Greaves, \& Coulson}]{dent05}
Dent, W.~R.~F., Greaves, J.~S., \& Coulson, I.~M. 2005, MNRAS, 359, 663

\bibitem[{{Draine}(1978)}]{draine78}
{Draine}, B.~T. 1978, \apjs, 36, 595

\bibitem[{Dullemond \& Dominik(2004)}]{duldom04}
Dullemond, C.~P. \& Dominik, C. 2004, \aap, 417, 159

\bibitem[{Dullemond {et~al.}(2002)Dullemond, {van Zadelhoff}, \&
  Natta}]{dullem02}
Dullemond, C.~P., {van Zadelhoff}, G.-J., \& Natta, A. 2002, \aap, 389, 464

\bibitem[{Dutrey {et~al.}(1997)Dutrey, Guilloteau, \& Gu\'elin}]{dutrey97}
Dutrey, A., Guilloteau, S., \& Gu\'elin, M. 1997, \aap, 317, L55

\bibitem[{{Gorti} \& {Hollenbach}(2004)}]{gorhol04}
{Gorti}, U. \& {Hollenbach}, D. 2004, \apj, 613, 424

\bibitem[{Habart {et~al.}(2004)Habart, Natta, \& Kr\"ugel}]{habart04}
Habart, E., Natta, A., \& Kr\"ugel, E. 2004, \aap, 427, 179

\bibitem[{Hauschildt {et~al.}(1999)Hauschildt, Allard, \& Baron}]{hausch99}
Hauschildt, P.~H., Allard, F., \& Baron, E. 1999, \apj, 512, 377

\bibitem[{Hogerheijde \& {van der Tak}(2000)}]{hogtak00}
Hogerheijde, M.~R. \& {van der Tak}, F. F.~S. 2000, \aap, 362, 697

\bibitem[{Jansen {et~al.}(1995)Jansen, van Dishoeck, Black, Spaans, \&
  Sosin}]{jansen95}
Jansen, D.~J., van Dishoeck, E.~F., Black, J.~H., Spaans, M., \& Sosin, C.
  1995, \aap, 302, 223

\bibitem[{Jonkheid {et~al.}(2004)Jonkheid, Faas, {van Zadelhoff}, \& {van
  Dishoeck}}]{jonkhe04}
Jonkheid, B., Faas, F. G.~A., {van Zadelhoff}, G.-J., \& {van Dishoeck}, E.~F.
  2004, \aap, 428, 511

\bibitem[{Jonkheid {et~al.}(2006)Jonkheid, Kamp, Augereau, \& {van
  Dishoeck}}]{jonkhe06}
Jonkheid, B., Kamp, I., Augereau, J.-C., \& {van Dishoeck}, E.~F. 2006, \aap,
  in press

\bibitem[{Kamp \& Dullemond(2004)}]{kamdul04}
Kamp, I. \& Dullemond, C.~P. 2004, \apj, 615, 991

\bibitem[{Kamp {et~al.}(2006)Kamp, Dullemond, Hogerheijde, \&
  Enriquez}]{kamp06}
Kamp, I., Dullemond, C.~P., Hogerheijde, M., \& Enriquez, J.~E. 2006, in
  Proceedings IAU symposiom 231, Astrochemistry: Recent Successes and Current
  Challenges, ed. D.~C. Lis, G.~A. Blake, \& E.~Herbst (Cambridge: Cambridge
  University Press), 377--386

\bibitem[{Kamp \& {van Zadelhoff}(2001)}]{kamzad01}
Kamp, I. \& {van Zadelhoff}, G.-J. 2001, \aap, 373, 641

\bibitem[{Kamp {et~al.}(2003)Kamp, {van Zadelhoff}, {van Dishoeck}, \&
  Stark}]{kamp03}
Kamp, I., {van Zadelhoff}, G.-J., {van Dishoeck}, E.~F., \& Stark, R. 2003,
  \aap, 397, 1129

\bibitem[{Kenyon \& Hartmann(1987)}]{kenhar87}
Kenyon, S.~J. \& Hartmann, L. 1987, \apj, 323, 714

\bibitem[{{Koerner} \& {Sargent}(1995)}]{koesar95}
{Koerner}, D.~W. \& {Sargent}, A.~I. 1995, \aj, 109, 2138

\bibitem[{{Nomura} \& {Millar}(2005)}]{nommil05}
{Nomura}, H. \& {Millar}, T.~J. 2005, \aap, 438, 923

\bibitem[{Spaans {et~al.}(1994)Spaans, Tielens, {van Dishoeck}, \&
  Bakes}]{spaans94}
Spaans, M., Tielens, A. G. G.~M., {van Dishoeck}, E.~F., \& Bakes, E. L.~O.
  1994, \apj, 437, 270

\bibitem[{{Thi} {et~al.}(2001){Thi}, {van Dishoeck}, {Blake}, {van Zadelhoff},
  {Horn}, {Becklin}, {Mannings}, {Sargent}, {van den Ancker}, {Natta}, \&
  {Kessler}}]{thi01}
{Thi}, W.~F., {van Dishoeck}, E.~F., {Blake}, G.~A., {et~al.} 2001, \apj, 561,
  1074

\bibitem[{Thi {et~al.}(2004)Thi, {van Zadelhoff}, \& {van Dishoeck}}]{thi04}
Thi, W.-F., {van Zadelhoff}, G.-J., \& {van Dishoeck}, E.~F. 2004, \aap, 425,
  955

\bibitem[{Tielens \& Hollenbach(1985)}]{tiehol85}
Tielens, A. G. G.~M. \& Hollenbach, D.~J. 1985, \apj, 291, 722

\bibitem[{{van Dishoeck}(1988)}]{dishoe88}
{van Dishoeck}, E.~F. 1988, in ASSL Vol. 146: Rate Coefficients in
  Astrochemistry, ed. T.~J. {Millar} \& D.~A. {Williams} (Dordrecht: Kluwer
  Academic Publishers), 49--72

\bibitem[{{van Dishoeck} \& Black(1982)}]{disbla82}
{van Dishoeck}, E.~F. \& Black, J.~H. 1982, \apj, 258, 533

\bibitem[{{van Zadelhoff} {et~al.}(2003){van Zadelhoff}, Aikawa, Hogerheijde,
  \& {van Dishoeck}}]{zadelh03}
{van Zadelhoff}, G.-J., Aikawa, Y., Hogerheijde, M., \& {van Dishoeck}, E.~F.
  2003, \aap, 397, 789

\end{thebibliography}

\end{document}